\documentclass[useAMS,usenatbib]{mn2e}
\usepackage{graphicx}
\usepackage{natbib}

\title[Constraints on sterile neutrino DM from \textit{XMM--Newton} observation of M33]
{Constraints on sterile neutrino dark matter from \textit{XMM--Newton} observation of M33}

\author[E.~Borriello et al.]{
E.~Borriello$^{1,2,3}$\thanks{enrico.borriello@desy.de},
M.~Paolillo$^{1}$, 
G.~Miele$^{1,2}$, 
G.~Longo$^{1,4}$, 
R.~Owen$^{5}$\\
$^1$Universit\`a “Federico II”, 
    Dipartimento di Scienze Fisiche, Via Cintia, Napoli, Italy\\
$^2$INFN Sezione di Napoli, 
    Via Cintia, Napoli, Italy\\
$^3$II. Institut f\"{u}r Theoretische Physik, Universit\"{a}t Hamburg, Germany\\
$^4$Visiting associate California Institute of Technology, Pasadena, 91125 Ca, USA\\
$^5$X-ray \& Observational Astronomy Group, 
    Dep. of Physics \& Astronomy, Univ. of Leicester, Univ. Road, Leicester LE1 7RH, U.K.
}

\begin{document}

\maketitle

\begin{abstract}

Using archival \textit{XMM--Newton} observations of the diffuse and unresolved 
components emission in the inner disc of M33 we exclude the possible contribution 
from narrow line emission in the energy range $0.5\div5$ keV more intense than 
$10^{-6}\div10^{-5}$ erg/s. Under the hypothesis that sterile neutrinos constitute 
the majority of the dark matter in M33, we use this result in order to put 
constraints on their parameter space in the $1\div10$ keV mass range.

\end{abstract}


\section{Introduction}

One of the main problems of modern astrophysics and cosmology is the unknown nature
of the dark matter (DM). Despite the fact that it represents the 80\% of the matter 
content of the Universe \citep{Komatsu:2010fb}, it has so far escaped any 
unambiguous direct detection attempt in underground laboratories. This has given 
rise to the development of several detection techniques aimed at finding signatures 
of DM annihilation or decay in the spectrum of cosmic rays \citep{Porter:2011nv}.

On the theoretical side, to solve this puzzle a wide range of possible solutions 
assuming new physics has been proposed. Standard particles are in fact unable to 
explain the evidence for the DM, while e.g., Supersymmetric and Kaluza-Klein 
extension of the Standard Model are known to provide a rich set of possible DM 
candidates in terms of weakly interacting massive particles. They have represented 
the standard assumption about the nature of DM for a long time, and even nowadays 
they constitute the preferred and most studied candidates in the literature. 
Nonetheless, many alternative solutions have been proposed. (See 
\citealt{Feng:2010gw} for an up to date review.)

Among these, a particularly interesting scenario assumes that DM is instead made 
up by $\mathcal{O}(1)$ keV sterile neutrinos \citep{Dodelson:1993je}. The well 
established discovery of neutrino masses points toward the possible existence of 
gauge singlet fermions responsible for the neutrino mass generation through a 
seesaw mechanism. The corresponding masses can range over all known mass scales, 
but if some of the singlet fermions are light, then sterile neutrinos emerge in 
the low-energy effective limit of the theory.  Such particles are predicted to 
decay into a lighter neutrino and a photon \citep{Pal:1981rm}, producing a narrow 
line in the X-ray spectrum at an energy equal to half the mass of the decaying 
neutrino. Such emission is expected to affect the formation of the first stars 
\citep{Biermann:2006bu} and to be directly detectable by existing and future 
X-ray telescopes. Sterile neutrinos could also explain the observed velocities of 
pulsars thanks to their anisotropic emission from a cooling neutron star born 
during the explosion of a supernova. (See \citealp{Kusenko:2009up} for a 
dedicated review.)

Searches for sterile neutrinos have been conducted both in dwarf and giant 
galaxies, using all available X-ray observatories.  For instance 
\citet{Loewenstein2009} fail to find signatures of sterile neutrino decay, 
using the Suzaku observations of the Ursa Minor dwarf spheroidal galaxy, and 
placed constraints on the existence of sterile neutrinos with given parameters. 
However, using the \textit{Chandra} X-ray observatory, \citet{Loewenstein:2009cm} 
reported the detection of a narrow emission feature with energy of 2.5 keV in the 
dwarf spheroidal (dSph) Willman 1, consistent with emission line from sterile 
neutrino radiative decay. However using the same data \citet{Mirabal:2010jj}, 
claim that all lines in the X-ray spectrum can be explained by residual background  
emission and spurious features arising from an incorrect modeling of the latter.
On the other hand \citet{Boyarsky:2010ci} analyze a combined sample of XMM 
observations of M31, Willman 1 and Fornax dSph as well as \textit{Chandra} data 
for Sculptor dSph, finding no evidence of a spectral feature at 2.5 keV, 
and confirming the previous findings of \citealp{Watson:2006qb}. (See, 
however, \citealp{Kusenko:2010sq} for a discussion of systematics.)
\citet{Mirabal2010} further used archival Swift XRT observation of the Milky Way 
satellite Segue 1, an ultra-faint dwarf galaxy with extreme mass-to-light ratio 
($M/L>1300$, see \citealp{Simon:2010ek} and \citealp{Geha:2008zr}). They find no 
evidence of emission lines due to neutrino decay either and put upper limits on 
its parameter space.

In this work we make use of archival \textit{XMM--Newton} observation of the 
diffuse and unresolved components emission of the inner disc of M33 
\citep{Owen:2009cg}. We set constraints on the sterile neutrino parameter space 
and compare them with analogous results previously found in the literature.

The paper is organized as follows: in section 2 we summarize the data set used in our analysis and the results we get; in section 3 we explore the consequences of our null detection in terms of bounds on the sterile neutrino parameter space; in section 4 we set our conclusions.

\begin{figure}
\centering
\includegraphics[angle=270,width=\columnwidth]{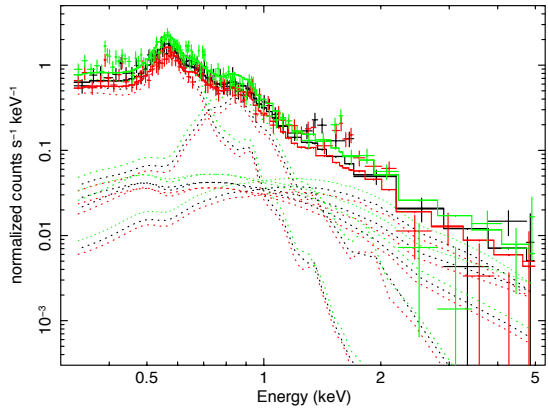}
\caption{The X-ray spectrum of the M33 inner region, from the PN camera; the best-fit model discussed in the text is overlayed, together with the individual components of the model. Note that only the two thermal components (those peaking at $\sim 0.6$ and $\sim 0.9$ keV) are free to vary, while all the others (due to the bright source contribution) are fixed to the residual flux expected to leak outside the masked region.}
\label{fit}
\end{figure}

\section{X-ray data analysis}
The X-ray data used in this work are those already discussed extensively by \citet{Owen:2009cg}, and summarized here for completeness. The original observations were part of the \citet{Pietsch2004} survey of M33. Among the entire dataset covering M33, we used only the observations \#0102640101, \#0102642301 and \#0141980801 from the EPIC-PN detector, i.e. the same used by \citet{Owen:2009cg} for their spectral analysis (see their Section 2.3) and covering the central part of the M33 disk out to a distance of $\sim 3.5$ kpc, with observing times ranging from 7 to 10 ks. The details of the X-ray data reduction (temporal filtering, event screening, etc.), can be found in \citealp{Owen:2009cg}.
We wish to emphasize here that the data were spatially filtered in order to remove the contamination due to bright sources down to a luminosity of $2\times 10^{35}$ erg s$^{-1}$. A residual contamination is estimated by the authors which, as done in the referred paper, is taken into account in the following spectral analysis.

\begin{figure}
\centering
\includegraphics[width=\columnwidth]{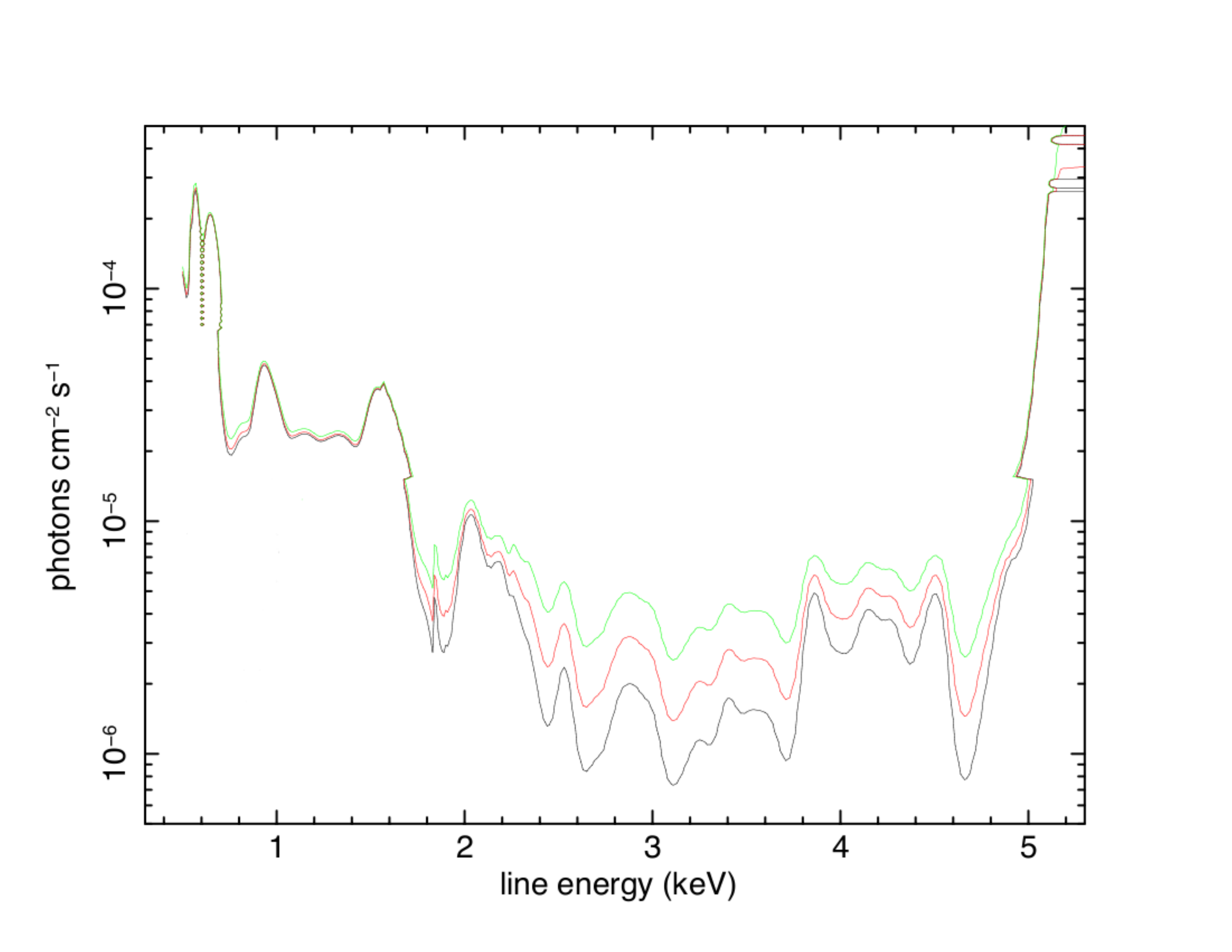}
\caption{68\%, 90\% and 99\% limits (blue, red and green line respectively) on the unresolved line normalization in the 0.5-5 keV energy interval.}
\label{limits}
\end{figure}

Appropriate Auxiliary Response File and Response Matrix File for the source-filtered region were created, while background files were produced from ``blank-sky'' fields extracted from a region of sky close to M33, and ``filter-wheel closed'' data to produce a background spectrum (again see details in \citealp{Owen:2009cg}). The final spectra for the three observations are shown in Figure \ref{fit}.

We first verified that we were able to reproduce the \citealp{Owen:2009cg} results, fitting (using the Cash statistics) the data with XSPEC 12.6 software\footnote{http://heasarc.nasa.gov/xanadu/xspec/} using a multi-component model consisting in two thermal plasma subject to Galactic absorption of $N_H=7.5\times 10^{20}$ cm$^{-2}$, constrained to solar abundance, with temperatures of $kT=0.2$ keV and $kT=0.6$ keV.
We also tested the use of models with subsolar ($Z_\odot = 0.1$) metallicity but the results are
consistent with those assuming solar metallicity.
 In addition a power law component was added to account for the 9\% flux from the bright sources leaking outside the masked regions; the bright ULX in the galaxy center, which can be fitted with a more complex spectrum (power law, disk black body, and intrinsic absorption) was treated separately, assuming that 3\% of its flux spills into our source-filtered region, as done in \citealp{Owen:2009cg}. 
We then added an unresolved line to our spectral model, and fitted iteratively the whole spectrum in the range 0.5-5 keV allowing the line flux, and the thermal components parameters (temperature and normalization), free to vary.

We computed the 68\%, 90\% and 99\% confidence levels for two interesting parameters (line flux and central energy), shown in Figure \ref{limits}, sampling our energy range in steps of 20 eV, in order to fully exploit the EPIC spectral resolution.
We note that several residual features are visible in the fitted spectrum, such as the emission lines around 1.5 keV due to Al-K$\alpha$ line. Furthermore, the spectral model we use does not account for unresolved point sources below the $2\times 10^{35}$ erg s$^{-1}$ threshold; this contribution is estimated by \citet{Owen:2009cg} to be $<1\%$ of the bright source emission. In order to minimize the uncertainties involved in refining the background spectrum, or in extrapolating the point source luminosity function to lower fluxes, we prefer to leave these contribution in the data, thus obtaining conservative lower limits on the DM line flux.

\begin{figure}
\centering
\includegraphics[width=\columnwidth]{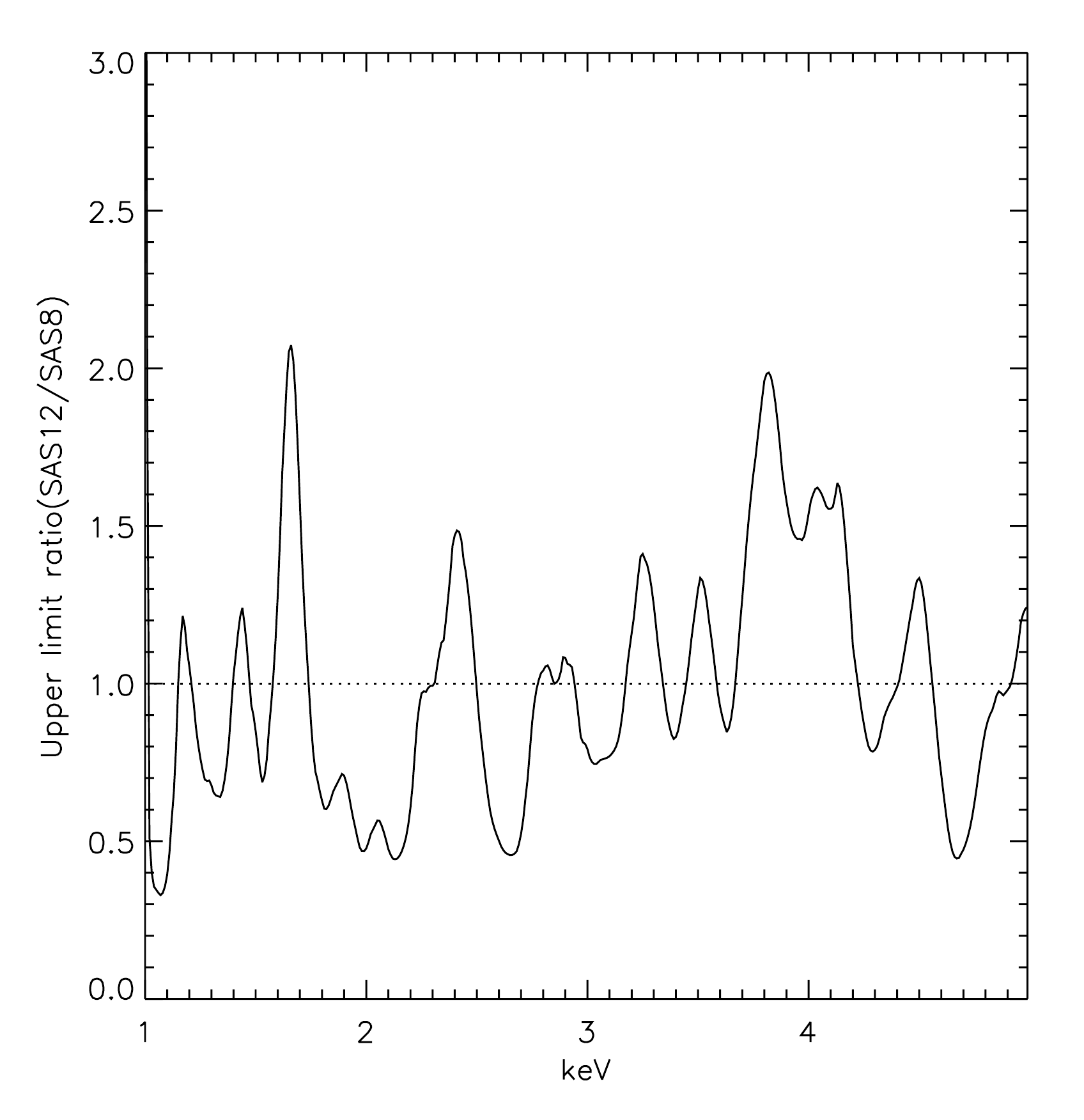}
\caption{Ratio between the flux upper limits derived with
the old (SAS v8) and new (SAS v11) calibration. Calibration issues introduce
at most a factor of two uncertainty in the result.}
\label{fig:conf}
\end{figure}

For consistency, as well as because of the very detailed checks performed in the \citet{Owen:2009cg} work, we prefer to use here the original data. Nonetheless, we checked the robustness of our results with respect to an update of the calibration software repeating the analysis --initially performed with SAS v8-- with the new SAS v11. The ratio between the flux upper limits derived with
the old and new calibration is plotted in Figure \ref{fig:conf}, and shows that calibration issues introduce an uncertainty, at most, of a factor of two in the result. We reasonably expect that further releases of the calibration software will not change the result more than this.

\section{X-ray flux from sterile neutrino dark matter}

We use the upper limits on the X-ray line flux derived in the previous section to constrain the possible contribution to M33 X-ray emission due to a radiative decay of sterile neutrino DM. This is done assuming that this additional contribution to the flux cannot exceed the upper limit deduced in the previuos section. 

Majorana sterile neutrinos may decay to a photon and an active neutrino \citep{Pal:1981rm}. The decay width of the process is\footnote{For Dirac sterile neutrinos the decay width of is half the one assumed here \citep{Pal:1981rm}.}
\[
 \Gamma(\nu_s\rightarrow\gamma\nu_a) = 
             1.36\times10^{-32}\frac{\sin^2 (2\theta)}{10^{-10}}\left(\frac{m_s}{\textrm{keV}}\right)^5 \textrm{s}^{-1}\ ,
\]
where $\theta$ is the neutrino mixing angle and $m_s$ --the double of the emitted photon energy-- is the mass of the sterile neutrino.

The resulting flux at the Solar System position is
 $\Phi = \Gamma\,\Omega\,S / 4\pi m_s$ ,
in which $\Omega$ and $S$ represent the angular size and the column density of the emitting region, respectively. In order to evaluate the DM column density, defined as the integral of the mass density along the line of sight, we need to parametrize the matter content of M33.
Since there is no universal consensus on the DM halo profile in M33, we tried both the cored and spiked profiles that are consistent with the galaxy rotation curve \citep{Corbelli2003}. The cored case is well represented by a
Burkert density profile \citep{Burkert1995}
\[
\rho(r) = \rho_0 (1+x)^{-1}(1+x^2)^{-1}, 
\]
while the spiked profile is well described by a NFW density
distribution \citep{Navarro:1996gj}
\[
\rho(r)=\rho_{0}x^{-1}(1+x)^{-2},
\]
with $x=r/r_0$. We consider two extremal cases among the various fits obtained in \citealp{Corbelli2003}, taking the $3\sigma$ lower limit for the DM density in the Burkert case, and the $3\sigma$ upper one in the NFW case. This way we are taking into account corresponding $3\sigma$ variation of the M/L ratio in the inner part of M33 \citep{Corbelli2003}.
Their defining parameters are:
 \begin{center}\begin{tabular}{|l|c|c|}\hline
 model   & $r_0$ (kpc)   & $\rho_0$ (GeV\,$c^{-2}$\,cm$^{-3}$)   \\ \hline
 NFW & 35       & 0.0765   \\
 Burkert & 12   & 0.336    \\ \hline
 \end{tabular}\end{center}
 Profiles with a central slope steeper than $r^{-1}$ like the Moore 
 profile \citep{Moore1999} are instead excluded \citep{Corbelli2003}. 

\begin{figure}
\centering
\includegraphics[width=\columnwidth]{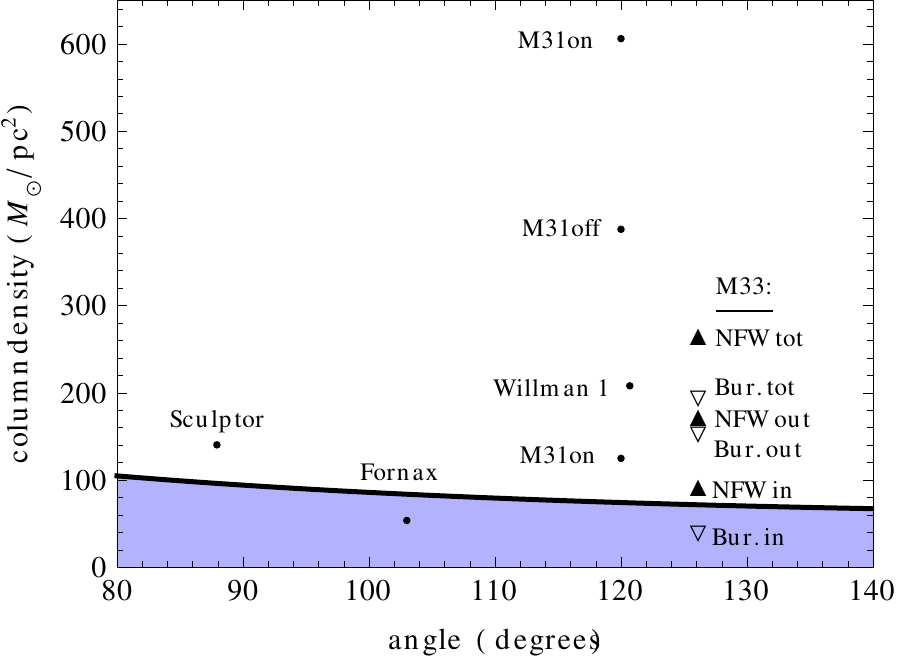}
\caption{Column density for both mass parametrization of M33 considered here compared to the ones assumed in previous similar analysis. The angle gives the angular distance from the galactic center ($126.1^{\circ}$ in the case of M33). The solid line represents the column density of the Milky Way (see text for more details).}
\label{fig:coldens}
\end{figure}

Figure \ref{fig:coldens} shows the DM column density for the various parametrization of M33 considered here: filled triangles in the NFW case, empty triangles for a Burkert profile. We assumed an infinite halo in calculating the mean surface density, because we checked that, repeting the calculation for a 17 kpc radius halo (the minimum possible radius, corresponding to the furthest observation in \citealp{Corbelli2003}) the result would change by a factor $\sim 1.3$. In each case we considered separately the inner region (in) of the galaxy --corresponding to the central 3.5 kpc bulge-- and the outer part (out) of it --defined as the cylindrical volume corresponding to the same angular region  and from which the central bulge has been subtracted. In addition we also considered the sum of both contributions (tot). In all the cases we took into account only the flux coming from the elliptical region analyzed in \citet{Owen:2009cg}. 
The values we obtain for the column density range from 40.4 to 266 $M_{\odot}/$pc$^2$. This precaution addresses the concerns raised by \citet{Kusenko:2010sq} about the uncertainties affecting the DM profile in the central region of the galaxy (see next section).
A comparison is made with the column density of the Milky Way and the other systems considered in \citealp{Loewenstein:2009cm} and \citealp{Boyarsky:2010ci}. 
In evaluating the total flux at the Solar System position we added up the contribution due to the Milky Way along the M33 line of sight, evaluated according to the DM density parametrization given in \citealp{Strigari:2007at}.

\begin{figure}
\centering
\includegraphics[width=\columnwidth]{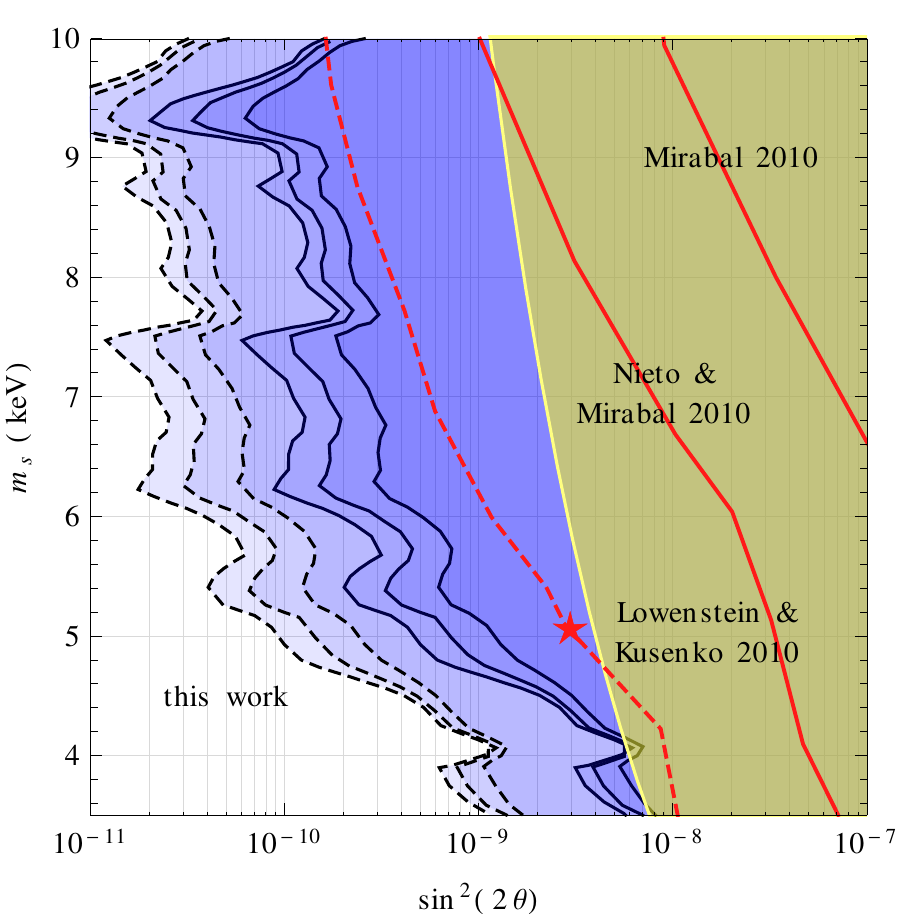}
\caption{Sterile neutrino parameter space. The light blue regions are excluded by X-ray observations of M33 (limits at 68\%, 90\% and 99\%).
The least (Bur. in, solid line) and the most (NFW tot, dotted line) stringent bounds are represented.
The red thick solid lines represent the constraints on sterile neutrinos derived for Segue 1 \citep{Mirabal2010} and Willman 1 \citep{Mirabal:2010jj}.
The red thin dashed line corresponds to the upper bound in Willman 1 \citep{Loewenstein:2009cm}. 
The star represents the model of \citet{Loewenstein:2009cm}.
The yellow line gives the models in which the cosmological amount of DM is entirely made up by sterile neutrinos. The yellow region is excluded by over-closure. The exclusion regions are deduced assuming 100\% of DM is made up of sterile neutrinos.}
\label{fig:exc}
\end{figure}

Figure \ref{fig:exc} shows the regions of the parameter space $(\sin^2 2\theta,m_s)$ excluded by the X-ray observations of M33. The main assumption here is that the DM halo is entirely made up by sterile neutrinos. Limits are shown at 1, 2, and 3 $\sigma$ confidence level. We show the most and the least stringent cases, among the six we considered here: NFW tot (dotted lines) and Bur. in (solid lines), respectively. The yellow line gives the models in which the cosmological amount of DM is entirely made up by sterile neutrinos. The yellow region is therefore excluded by over-closure. The star represents the model of \citet{Loewenstein:2009cm} which is found to be inconsistent with the observed X-ray emission from M33.

\section{Discussion and conclusions}

Let us now summarise the assumptions under which our results were obtained and the uncertainty affecting them. 
As pointed out in \citealp{Kusenko:2010sq}, the interaction of DM with baryons could have the effect of expelling the DM out of the central part of a spiral galaxy \citep{Klypin:2001xu}. DM could therefore provide only a sub-dominant contribution to the total amount of mass in the central part of M33. This is the reason why we separated the emission along the line of sight in two contributions, the one produced in the innermost region of M33, and the contribution coming from the two external cylinders. Interestingly (see Figure \ref{fig:coldens}), these external (out) contributions --whose mass distribution is determined on a more stable basis-- are the ones giving the most stringent bounds. This makes our results stable against the main criticism raised by \citet{Kusenko:2010sq}. We stress that our results change by less than one order of magnitude because of the uncertainty affecting the DM density distribution, that could be either cored or spiked at the center of M33.

An additional uncertainty factor is represented by the effect of photometric
absorption. This will reduce the detected X-ray flux, thus resulting in an
underestimate of the M33 DM emission. We note however that: i) galactic absorption
has already been taken into account in the spectral fitting described in Sec.2; ii)
photometric absorption is relevant only in the soft X-ray band below 2 keV, and thus
the limits in the hard energy range are essentially unaffected; iii) additional
intrinsic M33 absorption may be present but likely with a low covering factor. The
population of M33 X-ray sources for instance (see e.g. discussion
in \citealp{Foschini2004} and \citealp{Grimm2007} does not reveal widespread absorption, and anyway with
$n_H$ less than a few $10^{21}$ cm$^{-2}$ which results in a factor of $\sim 2$
reduced flux at 1 keV. Nevertheless, to evaluate the effect that absorption might have
on our estimates, we could simply divide by $\sim2$
 the emission coming from the back cilinder. This would result in a 25\,\% reduction of the total flux (out cases), 
which does not affect our main results.
We also need to point out that in our analysis we did not try to fine tune the
background removal, nor we did model the contribution of the unresolved X-ray binary
population, in order to avoid biases due to the poor knowledge of these
contributions to the overall X-ray emission. This choice results in conservative
upper limits on the M33 DM content, since the estimated emission is actually
overestimated.

In summary, we make use of archival \textit{XMM--Newton} observation of the diffuse and unresolved components emission of the inner disc of M33 \citep{Owen:2009cg}, finding no evidence for line emission compatible with sterile neutrino radiative decay. We accordingly set bounds on sterile neutrino parameter space under the assumption that they account for the entire DM content of M33. In particular we have found no evidence for the emission by 5 keV neutrinos deduced by \citet{Loewenstein:2009cm}.

\section*{Acknowledgments}

EB acknowledges financial support by the Deutsche Forschungsgemeinschaft through SFB 676 ``Particles, Strings and the Early Universe: The Structure of Matter and Space-Time.'' EB and GM acknowledge support by the INFN I.S. FA51 and the PRIN 2008``Fisica Astroparticellare: Neutrini ed Universo Primordiale'' of the MIUR. MP acknowledges support from PRIN 2009 of the MIUR. GL thanks the California Institute of Technology for hospitality.

\end{document}